\documentclass[aps,prf,onecolumn,10pt]{revtex4-2}
\usepackage{amsmath}
\usepackage{graphicx}
\usepackage{xcolor}
\usepackage{hyperref}
\usepackage{amssymb}

\usepackage{geometry}
\newcommand{\modif}[1]{\textcolor{black}{#1}}

\begin{document}

\title{\modif{Kelvin-Helmholtz} instability and formation of viscous solitons on highly viscous liquids}

\author{M. Aulnette}
\author{J. Zhang}
\author{M. Rabaud}
\author{F. Moisy}
\email{frederic.moisy@universite-paris-saclay.fr}

\affiliation{Universit\'e Paris-Saclay, CNRS, FAST, 91405, Orsay, France.}

\vspace{0.4 cm}

\date{\today}

\begin{abstract}
\vspace{0.8 cm}

\parbox{12.2cm}{
Viscous solitons are strongly non-linear surface deformations generated by blowing wind over a liquid beyond a critical viscosity. Their shape and dynamics result from a balance between wind drag, surface tension and viscous dissipation in the liquid. We investigate here the influence of the liquid viscosity in their generation and propagation. Experiments are carried out using silicon oils, covering a wide range of kinematic viscosities $\nu_\ell$ between 20 and 5000~mm$^2$~s$^{-1}$. \modif{We show that, for $\nu_\ell> 200$~mm$^2$~s$^{-1}$, viscous solitons are sub-critically generated from an unstable initial wave train at small fetch, where the wind shear stress is larger. The properties of this initial wave train are those expected from Miles's theory of the Kelvin-Helmholtz instability of a highly viscous fluid sheared by a turbulent wind: the critical friction velocity and critical wavelength are independent of $\nu_\ell$, and the phase velocity decreases as $\nu_\ell^{-1}$.} We demonstrate the subcritical nature of the transition to viscous solitons by triggering them using a wavemaker for a wind velocity below the natural threshold. Finally, we analyze the flow field induced by a viscous soliton, and show that it is well described by a two-dimensional Stokeslet singularity in the far field. The resulting viscous drag implies a propagation velocity with a logarithmic correction in liquid depth, in good agreement with our measurements.
}

\end{abstract}

\maketitle

\vspace{0.6 cm}

%---------------------------------------------------------------------------------------------------------------------
\section{Introduction} \label{sec:intro}

When wind blows over a highly viscous liquid, localized liquid bumps, of height of the order of the capillary length, are generated and propagate downwind~\cite{Francis_1954,francis1956lxix,andritsos1987interfacial,andritsos1989effect,Paquier_2016,Aulnette_2019}. These strongly nonlinear coherent structures, called ``viscous solitons'', contrast with the weakly nonlinear waves observed at low viscosity~\cite{lin2008direct,Paquier_2015,zavadsky2017two,shemer2019evolution,Wu_2021}. They are sub-critically generated from an unstable localized wave train, and propagate downwind at a velocity that results from a balance between forcing by wind forcing in the air and viscous dissipation in the liquid \cite{Aulnette_2019}. 

Viscous solitons were first observed by Francis \cite{Francis_1954,francis1956lxix} in a channel flow over a developing boundary layer. The initial wave train from which they arise was identified by Miles \cite{Miles1959generation,Miles1993surface} as the result of the Kelvin-Helmholtz instability~\cite{Rabaud_2020}. These structures were also observed by Andritsos {\it et.~al} \cite{andritsos1987interfacial,andritsos1989effect} in circular pipe flows, who named them ``irregular large-amplitude waves.'' Their linear stability analysis also suggests that they are triggered by the Kelvin-Helmholtz instability.

Recently, the dynamics of viscous solitons was analyzed using laser sheet profilometry and particle image velocimetry in silicon oil of kinematic viscosity $\nu_\ell = 1000$~mm$^2$~s$^{-1}$ \cite{Aulnette_2019}. This study highlighted their subcritical nature: since the wind shear stress naturally decreases with fetch in a developing boundary layer, viscous solitons are systematically generated at small fetch, where the shear stress is larger, but they can propagate downwind in regions of lower shear stress. 

An intriguing property of viscous solitons is that they are generated only for sufficiently large liquid viscosity, while lower viscosities produce classical weakly nonlinear waves.  Stronger non-linearities favored by larger viscosity is an uncommon situation, in apparent contradiction with most conventional hydrodynamic instabilities. A qualitative explanation for this behavior is that at large viscosity waves become over-damped~\cite{Lamb,Leblond87}: the energy propagation velocity vanishes, so  the energy supplied by the wind cannot be radiated by the waves, resulting in a strong {\it in situ} amplification of any disturbance. As the deformation amplitude becomes large, the wind drag strongly increases, and the wave crest becomes directly pushed by the wind: the wave energy focuses on a single object, a ``particle,'' with a velocity no longer set by the restoring forces of the wave (gravity and surface tension), but by a balance between air drag and viscous dissipation.

This transition from waves to viscous solitons in a highly viscous liquid raises a number of questions:  What is the minimum viscosity beyond which solitons can be generated? How sharp is the transition from weakly nonlinear waves to viscous solitons?  How do the viscous solitons' properties (shape, velocity) depend on viscosity? May viscous solitons be artificially produced from a disturbance other than an unstable wave train? 

The nature of the unstable wave from which viscous solitons arise was first analyzed by Miles~\cite{Miles1959generation}. Analyzing the data of Francis  \cite{Francis_1954,francis1956lxix}, he extended the original inviscid Kelvin-Helmholtz framework by including liquid viscosity and sheared wind profile. He concludes that, unlike wind-generated waves over water or liquids of low viscosity, the Kelvin-Helmholtz mechanism governs the initial instability at large liquid viscosity. \modif{The key results is a critical friction velocity and critical wavelength independent of the liquid viscosity~\cite{Miles1959generation}. In a refined analysis, he includes the effects of the turbulent stresses in the air, and predicts a phase velocity decreasing as $\nu_\ell^{-1}$~\cite{Miles1993surface}.}

\modif{In the present paper, we address Miles's predictions by systematically varying the liquid viscosity over a wide range ($\nu_\ell = 20-5000$~m$^2$~s$^{-1}$) with all other fluid properties kept constant (fluid density, surface tension, turbulent boundary layer thickness etc.) We confirm that, for sufficiently large viscosity, the critical friction velocity and critical wavelength are independent of $\nu_\ell$, while the phase velocity decreases as $\nu_\ell^{-1}$. We also propose a simplified version of Miles's stability argument that provides a correct description of the measured wavelength.}

Another issue is the subcritical nature of the Kelvin-Helmholtz instability that leads to viscous solitons. In the experiments of Aulnette {\it et al.} \cite{Aulnette_2019}, the soliton branch of the bifurcation diagram was reconstructed using the natural decrease of the wind shear stress with fetch in a developing boundary layer. However, the unstable branch of the diagram in the hysteresis window remains inaccessible by this method. Here, using an immersed wavemaker to mechanically excite solitons, we confirm that they can be generated for a shear stress below the natural threshold. A similar approach has been used by Meignin {\it et al.}~\cite{Meignin:2003} to excite subcritical Kelvin-Helmholtz waves in a Hele-Shaw cell between parallel plates separated by a small gap.  Such a subcritical excitation is analogous to the take-off of a kite: if the wind is too low for a spontaneous take-off, a vertical impulse must be imparted for the kite to reach the altitude at which the wind force balances its weight.  From the minimum impulse necessary to generate a soliton for a given wind, we reconstruct here the unstable branch and complete the subcritical bifurcation diagram of viscous solitons.

%---------------------------------------------------------------------------------------------------------------------

\section{Experimental set-up and critical friction velocity} \label{sec:experiments}

\begin{figure}[tb]
\begin{center}
\includegraphics[width=15cm]{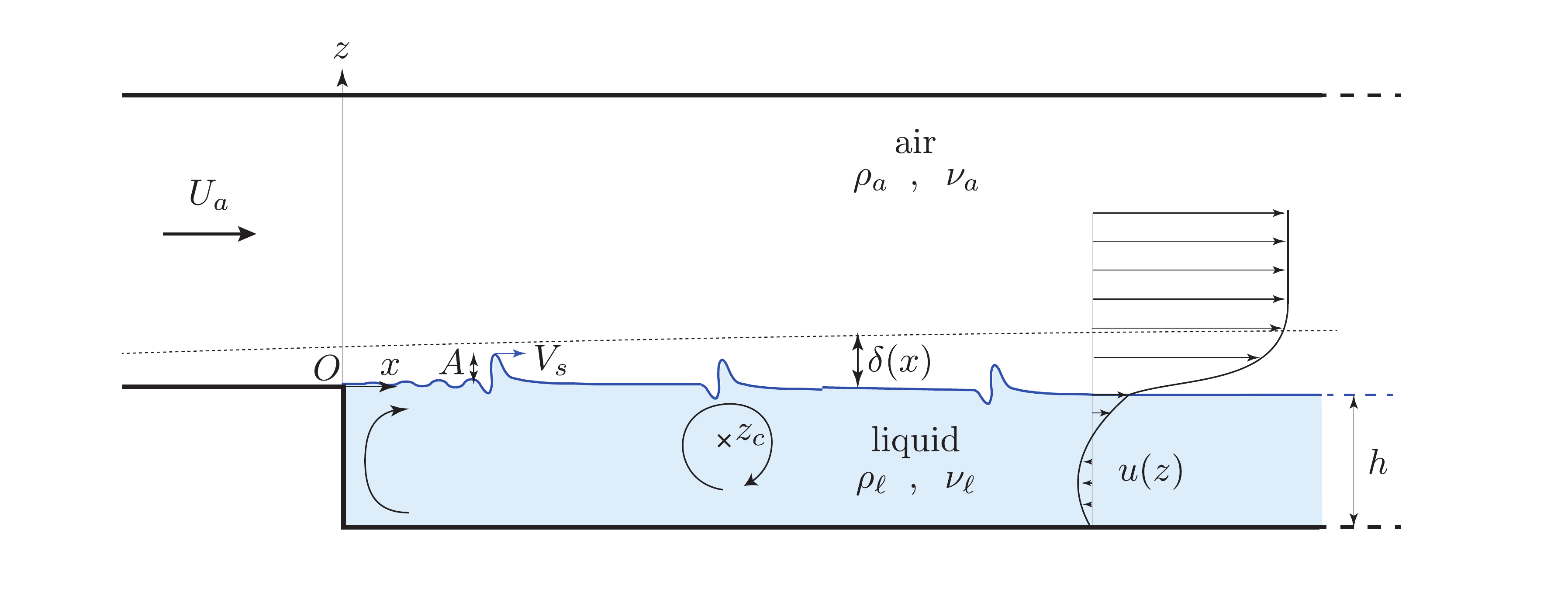}
\caption{Sketch of the experimental setup. Air is blown above a liquid bath of depth $h$. $A$ is the soliton amplitude, $V_s$ the soliton velocity, $z_c$ the depth of the vortex center under the soliton, and $\delta(x)$ the boundary layer thickness in the air. Velocity profiles in the air and the liquid are not drawn to scale,  and the soliton amplitude is magnified for visibility.}
\label{fig:setup}
\end{center}
\end{figure}

The experimental setup is sketched in Fig.~\ref{fig:setup} and is only briefly described here (more details can be found in Refs.~\cite{Paquier_2015,Aulnette_2019}). It is composed of a wind-tunnel, 105 mm high and 296 mm wide, at the bottom of which is fitted a liquid-filled rectangular tank. 
Two tanks are used: a long one (1.5 m) of maximum depth $h= 50$ mm, and a shorter one (0.5 m) of maximum depth $h=150$ mm. In each tank the liquid depth can be reduced by adding an immersed Plexiglas plate at the bottom, allowing us to vary the depth from 20 to 150 mm. The turbulent air flow, of mean velocity $U_a$ in the range $1-15$~m~s$^{-1}$, is blown in the wind tunnel through a convergent section located upstream.

The working fluid is silicon oil (Bluesil Fluid 47), of kinematic viscosity $\nu_{\ell}$ in the range 20~mm$^2$~s$^{-1}$ to 5000~mm$^2$~s$^{-1}$. Over this wide range of viscosity,  density and surface tension remain almost constant, $\rho_{\ell} \simeq 970 \pm 5$ kg~m$^{-3}$ and $\gamma \simeq 21.0 \pm 0.1$ mN~m$^{-1}$ at 25$^\mathrm{o}$. The capillary length and minimum phase velocity,
\begin{equation}
    \lambda_c = 2\pi \sqrt{\frac{\gamma}{\rho_{\ell} g}}, \qquad c_{\min} = \left( \frac{4 g \gamma}{\rho_\ell} \right)^{1/4}, 
    \label{eq:lc}
\end{equation}
are therefore constant, $\lambda_c = 9.4$~mm and $c_{\min} = 171$~mm~s$^{-1}$, for all viscosities.

Surface deformations are measured using laser sheet profilometry (LSP) \cite{Aulnette_2019,Aulnette_PhD_2021}:  we image the intersection between a tilted laser sheet and the liquid surface, made diffusive by adding to the oil a suspension of a white pigment (titanium oxide). We compute the surface elevation by a standard edge detection method, with a vertical resolution of 0.1~mm.  Measurements are performed along $x$ at a constant distance from one side wall, $y=50$~mm. The flow below the surface is measured using particle image velocimetry (PIV) in a vertical plane along $x$, with a spatial resolution of 0.7~mm. For these measurements the oil is seeded with small glass beads (10~$\mu$m diameter), without white pigment.

\begin{figure}[tb]
\begin{center}
\includegraphics[width=15cm]{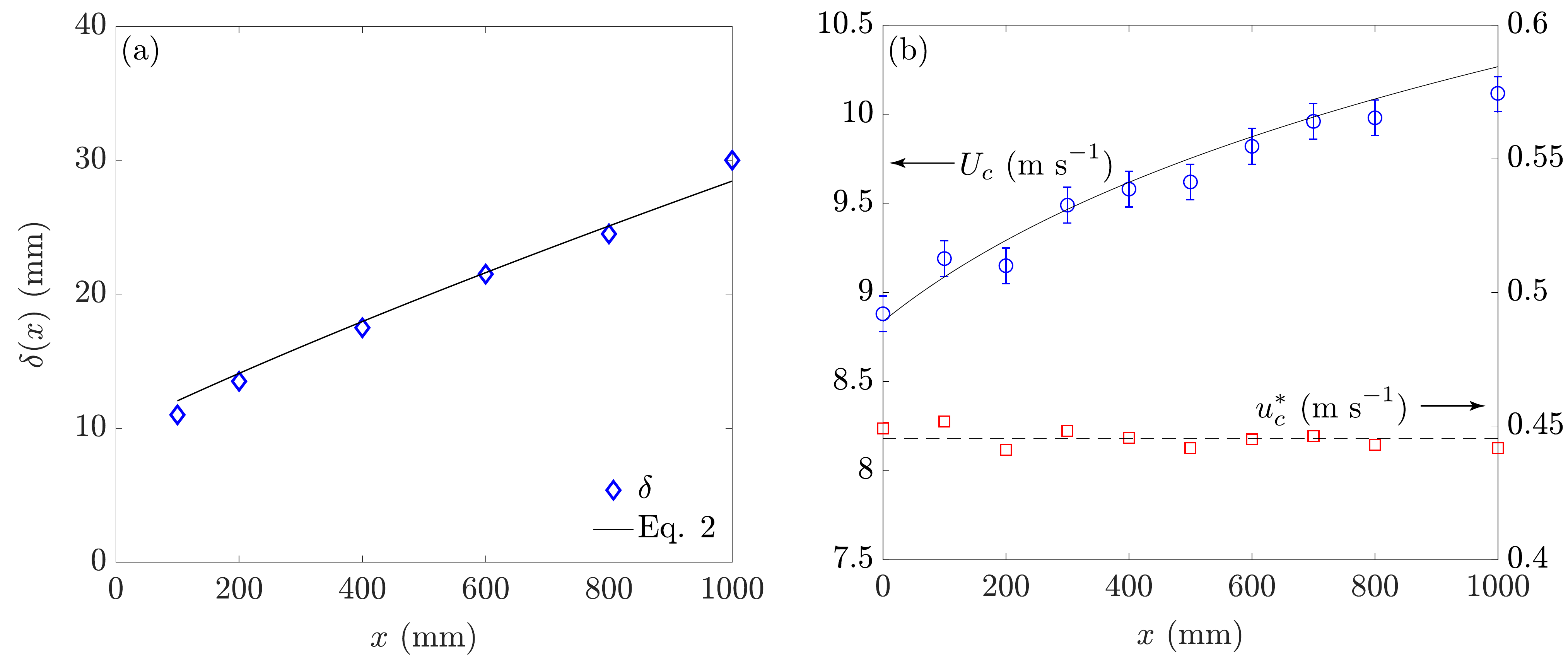}
\caption{(a) Boundary layer thickness $\delta(x)$ measured at the onset of viscous solitons as a function of the fetch $x$, defined such that $u(z=\delta) \simeq 0.99 U_c$. The line shows the best fit by Eq.~(\ref{Eq_delta_x}).
(b) Blue circles: critical wind velocity $U_c$ for the onset of viscous solitons as a function of the fetch $x$, as obtained by shifting the onset of solitons using a membrane of length $x$, for a viscosity $\nu_{\ell} = 1000$~mm$^2$~s$^{-1}$. Red squares: corresponding critical friction velocity $u^*_c$ deduced from Eq.~(\ref{Eq_tau_x})-(\ref{Eq_u*}). }
\label{fig:membrane}
\end{center}
\end{figure}

The flow in the air, characterized by hot-wire anemometry, is a turbulent boundary layer developing above the interface. The increase of the boundary layer thickness $\delta(x)$, plotted in Fig. 2(a), is well described by the classical law for a developing turbulent boundary layer over a flat rigid wall~\cite{Schlichtling}, 
\begin{equation}
   \delta(x) \simeq 0.37 (x + x_0) Re_x^{-0.2},
   \label{Eq_delta_x}
\end{equation}
with $Re_x = U_a (x + x_0) / \nu_a$ the Reynolds number based on the total streamwise distance from the beginning of the boundary layer; here $x_0=  350$~mm is the length of the flat plate between the end of the wind-tunnel convergent and the edge of the liquid tank at $x=0$ \cite{Aulnette_PhD_2021}. This thickening implies a decrease with $x$ of the shear stress $\tau(x)$ applied at the liquid surface, in the form
\begin{equation}
   \tau(x) \simeq C \rho_a U_a^2 Re_x^{-0.2}.
   \label{Eq_tau_x}
\end{equation}
In the following,  we use the local friction velocity
\begin{equation}
u^*(x) = \sqrt{\tau(x) /\rho_a}
\label{Eq_u*}
\end{equation}
as the relevant control parameter. We determine this quantity using the continuity of the shear stress at the interface, by measuring the induced surface current $U_s(x)$. Because of the large viscosity of the liquid, the flow in the liquid is laminar and, below the onset of waves, well described by the Stokes solution $u(x,z) \simeq U_s(x) (1+z/h)(1+3z/h)$ for $-h \leq z \leq 0$, with
\begin{equation}
U_s(x) = \frac{\tau(x) h}{4 \eta_{\ell}} = \frac{s u^{*2}(x) h}{4 \nu_\ell},
\label{eq:Us}
\end{equation}
with $s= \rho_a / \rho_l$ the density contrast and $\eta_\ell = \rho_l \nu_\ell$. For a liquid depth $h=50$~mm, this surface drift velocity ranges from 25~mm~s$^{-1}$ for $\nu_\ell = 100$~m$^2$~s$^{-1}$ down to 0.5~mm~s$^{-1}$ for $\nu_\ell = 5000$~m$^2$~s$^{-1}$, which is always negligible compared to the mean air velocity. The corresponding Reynolds number in the liquid, $U_s h/\nu_\ell$, is in the range $10^{-2}-10$, confirming that this mean flow is laminar. The best fit of the shear stress $\tau(x)$ deduced from $U_s(x)$ with Eq.~(\ref{Eq_tau_x}) gives $C\simeq 0.029$, a value similar to that obtained for a turbulent boundary layer over a flat rigid plate~\cite{Schlichtling}.

Another consequence of this wind shear stress is that the liquid surface is not perfectly horizontal but tilted by a small angle, of the order of 0.1$^\mathrm{o}$. This angle increases linearly with $\tau$ and does not depend on the liquid viscosity. We subtract this mean tilted surface profile from the height $\zeta(x,t)$ measured by LSP, and define the wave profile as $\xi(x,t) = \zeta(x,t) - \langle \zeta(x,t) \rangle_t$, where $\langle \cdot \rangle_t$ is the average over time.

The decrease of the shear stress along the liquid tank (typically $-30\%$) plays an important role in the generation and propagation of viscous solitons. Because of the decay of $u^*$ with $x$, viscous solitons are primarily generated at small fetch. It is possible however to shift their generation further downstream,  by covering the liquid between $0$ and $x$ by a light floating membrane of length $x$.  This is equivalent to extending the length of the rigid-wall boundary layer, and thus increases the thickness of the boundary layer at the beginning of the liquid surface. Figure~\ref{fig:membrane}(b) shows the critical wind velocity $U_c$ and friction velocity $u^*_c$ for the onset of viscous solitons for a liquid viscosity $\nu_\ell = 1000$~m$^2$~s$^{-1}$ for various locations $x$  (i.e. for various membrane lengths). This clearly shows that the minimum wind velocity $U_c$ to generate solitons increases with $x$, but that the friction velocity $u^*_c$ remains constant, $u_c^{*} \simeq 0.44 \pm 0.01$~m~s$^{-1}$ ($\tau = 0.24 \pm 0.01$~Pa), confirming that the shear stress governs the instability.

%---------------------------------------------------------------------------------------------------------------------
\section{Influence of viscosity on the dynamics of viscous solitons} \label{sec:onset_iwp}

\begin{figure}[tb]
\begin{center}
\includegraphics[width=10cm]{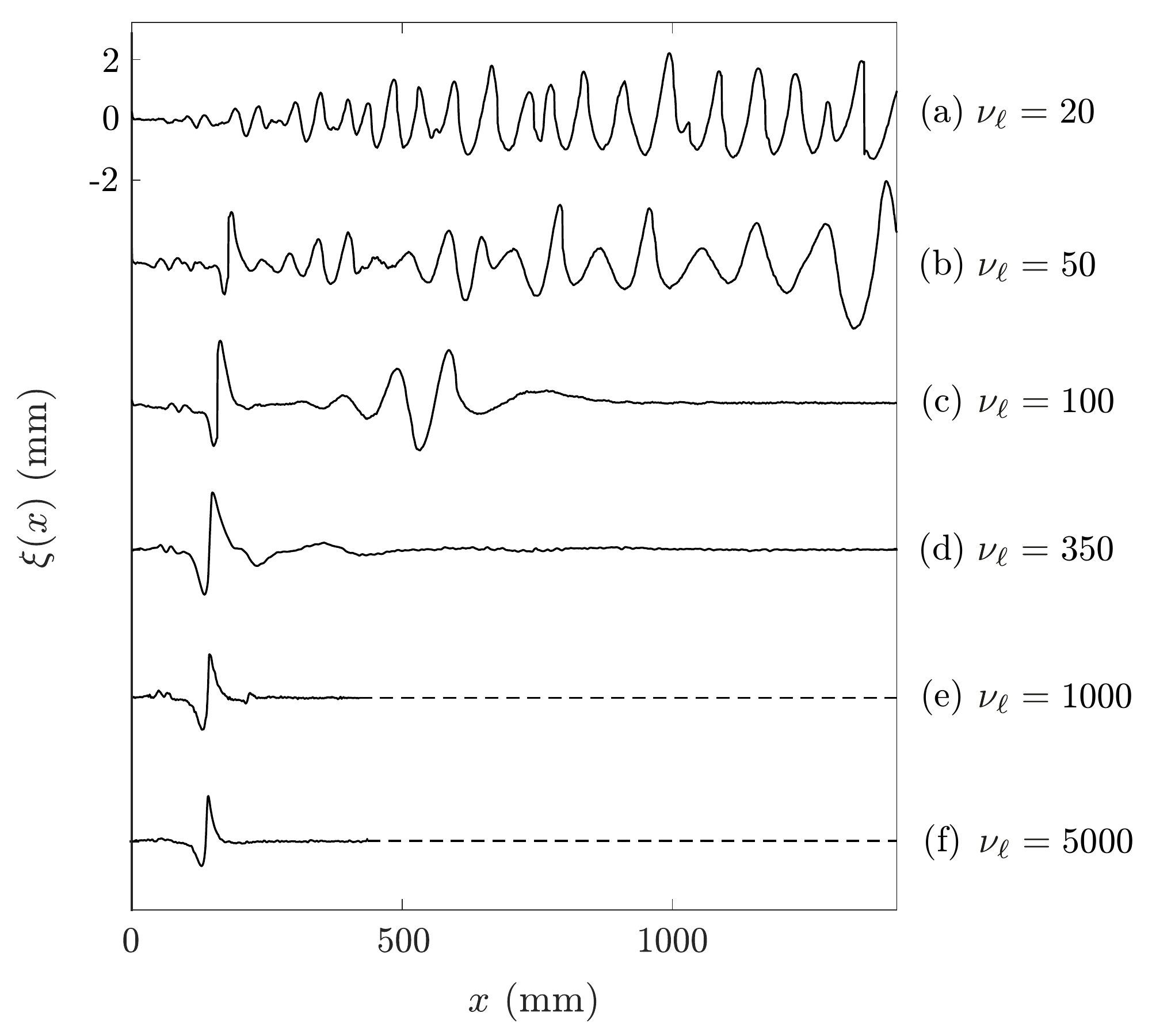}
\caption{\modif{Typical wave patterns measured by LSP for increasing liquid viscosities $\nu_\ell$, showing the gradual evolution between regular waves at small viscosity and viscous solitons at large viscosity (the kinematic viscosity $\nu_\ell$ is in mm$^2$~s$^{-1}$). In each case, the wind velocity $U_a$ is approximately 10\% larger than the onset of waves.  Measurements are restricted to $x<420$~mm in (e), (f).}}
\label{fig:profiles}
\end{center}
\end{figure}

%nu=V20         Ua=6.45 ms-1 seuil 4.79 (3.13 ?)  (1.35)
%nu=V50         Ua=7.29 ms-1 seuil 5.76  (1.26)
%nu=V100       Ua=7.93 ms-1  seuil 7.39  (1.07)
%nu=V350        Ua=9.2 ms-1   seyuil 8.75 (1.06)
%nu=V1000     Ua=9.08 ms-1   seuil 8.75  (1.04)
%nu=V5000     Ua=9.16 ms-1   seuil 8.75  (1.05)

\begin{figure}[tb]
\begin{center}
\includegraphics[width=15cm]{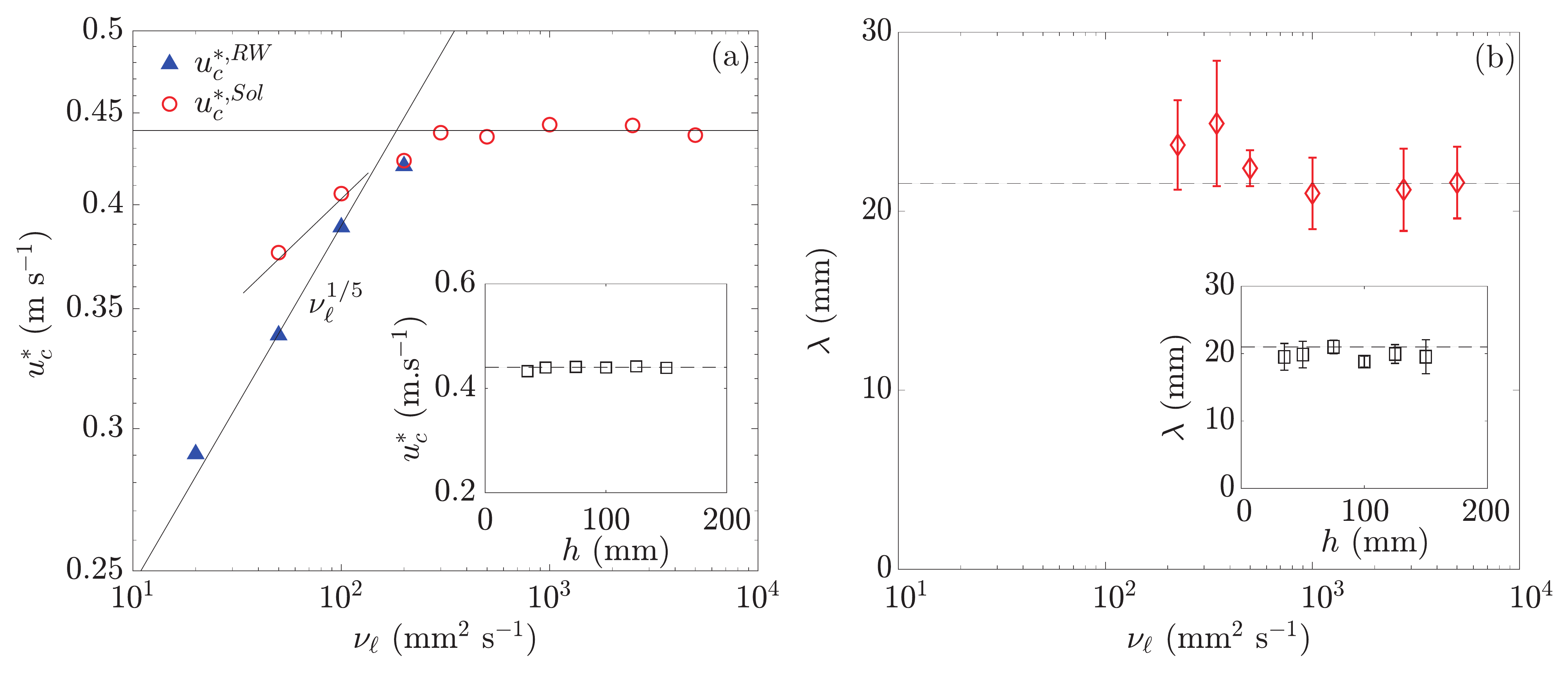}
\caption{(a) Critical friction velocity for the onset of regular waves ($u_c^{*RW}$, blue triangles) and viscous solitons ($u_c^{*Sol}$, red circles)  as a function of the liquid viscosity $\nu_{\ell}$ for a liquid depth $h=35$ mm. At large viscosity, both quantities merge to a constant value. (b) Critical wavelength $\lambda$ of the initial wave packet as a function of $\nu_{\ell}$ for $h=35$ mm. The insets in (a) and (b) show that these quantities are independent of the liquid depth $h$ (here for $\nu_{\ell} = 1000$~mm$^2$~s$^{-1}$).}
\label{fig:seuil}
\end{center}
\end{figure}

\modif{We now characterize the effect of the liquid viscosity on the wave dynamics. Figure~\ref{fig:profiles} illustrates the evolution of the wave patterns as the liquid viscosity is increased, for a wind velocity approximately 10\% larger than the onset of waves. At the lowest viscosity ($\nu_\ell =20$~mm$^2$~s$^{-1}$), the surface shows weakly nonlinear waves of amplitude slowly increasing with the fetch, similar to the waves found in classical air-water experiments. These regular waves are triggered when the friction velocity reaches a critical value, $u_c^{*RW}$ [blue triangles in Fig.~\ref{fig:seuil}(a)]. For larger viscosity ($\nu_\ell =50$~mm$^2$~s$^{-1}$), regular waves are still observed, with a critical friction velocity $u_c^{*RW}$ that increases as $\nu_\ell^{1/5}$. This scaling, in good agreement with the experiments of Paquier {\it et al.}~\cite{Paquier_2016}, can be interpreted as the transition from wrinkles to regular waves. Wrinkles are disorganized fluctuations of small amplitude (of the order of 1-10~$\mu$m, below the LSP resolution) excited by the pressure fluctuations traveling in the boundary layer~\cite{Perrard_2019,Nove_2020}.  They correspond to the inviscid growth mechanism of Phillips~\cite{phillips1957generation} saturated by the viscous dissipation in the liquid.}

\modif{In addition to this first transition, a second transition is observed at intermediate viscosities, for $\nu_\ell \geq 50$~mm$^2$~s$^{-1}$. For $u^* \geq u_c^{*Sol}$ [red circles in Fig.~\ref{fig:seuil}(a)], viscous solitons are emitted at small fetch, superimposed to the background of regular waves [Fig.~\ref{fig:profiles}(b)]. Solitons propagate over a certain distance until they become unstable and excite weakly nonlinear waves.   The thresholds for regular waves and viscous solitons, $u_c^{*RW}$ and $u_c^{*Sol}$, both increase with $\nu_\ell$ and  merge toward a constant value, $u_c^* \simeq 0.44 \pm 0.01$~m~s$^{-1}$  for $\nu_\ell \geq 350$~mm$^2$~s$^{-1}$. In that large-$\nu_\ell$ regime, viscous solitons have their characteristic shape, with an almost vertical rear facing the wind and a weak slope at the front. The constancy of the critical friction velocity is also checked with respect to the liquid depth for  $h = 20 - 150$ mm [see the inset in Fig.~\ref{fig:seuil}(a)].  }

\begin{figure}[tb]
\begin{center}
\includegraphics[width=15cm]{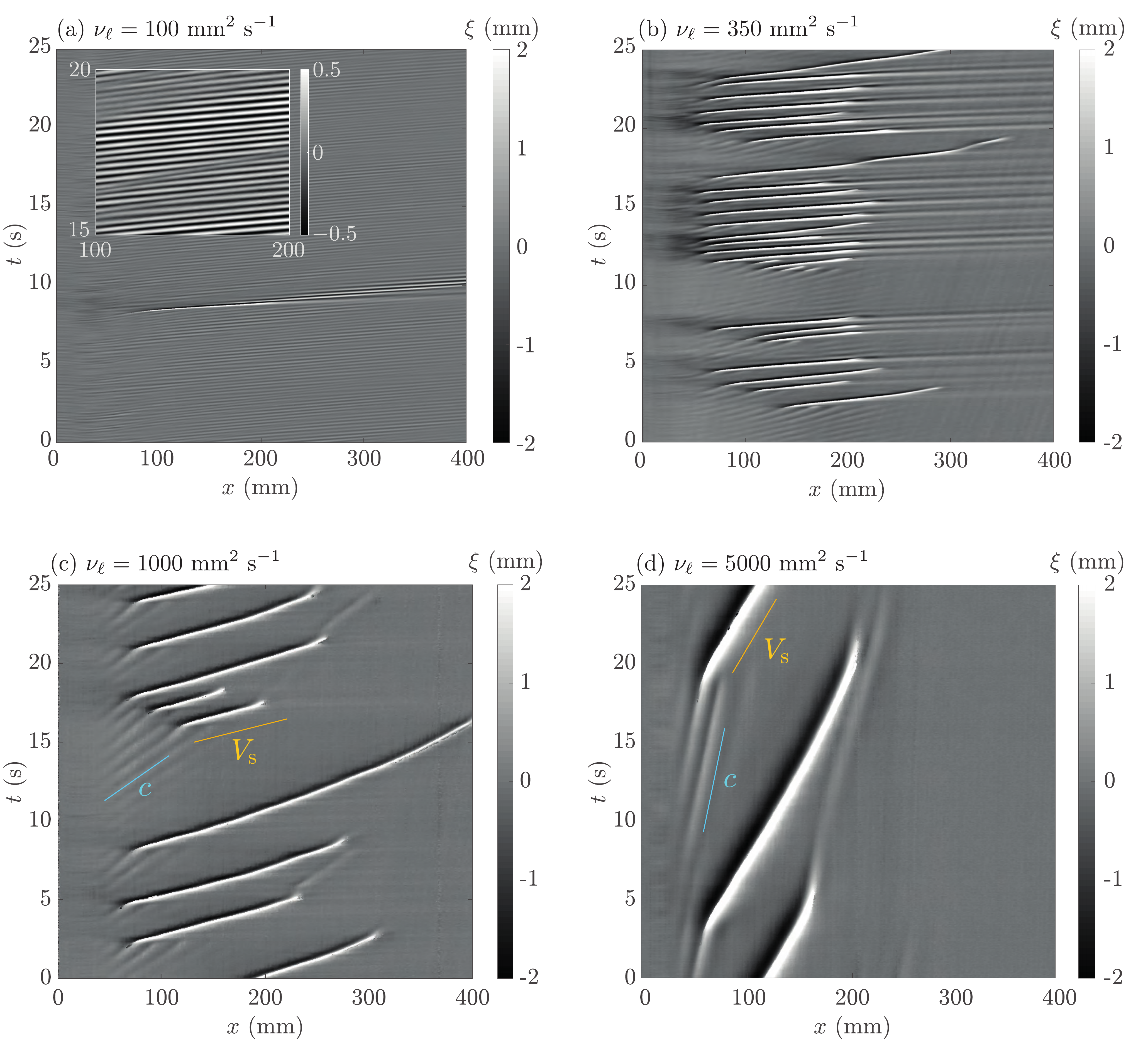}
\caption{Spatiotemporal diagrams of the surface elevation $\xi(x,t)$ during 25 s for a wind velocity slightly above the onset ($U_a\simeq 1.01 U_c$), for a liquid depth $h = 35$~mm and for four liquid viscosities: (a) $\nu_{\ell} = 100$~mm$^2$~s$^{-1}$, (b) $\nu_{\ell} = 350$~mm$^2$~s$^{-1}$, (c) $\nu_{\ell} = 1000$~mm$^2$~s$^{-1}$, and (d) $\nu_{\ell} = 5000$~mm$^2$~s$^{-1}$. At small viscosity (a), the surface is covered by regular waves at all $x$ (visible in the enhanced contrast rectangle) with erratic emission of strongly nonlinear wave packets (here at $t\simeq 8$~s). At larger viscosity (b)-(d), a localized wave packet is formed at small fetch, from which viscous solitons are emitted. \modif{Blue lines show the phase velocity $c$ of the initial wave packet, and orange lines the propagation velocity $V_s$ of viscous solitons.}}
\label{fig:spatios}
\end{center}
\end{figure}

\modif{We focus in the following on the viscous soliton regime at large viscosity.  Their dynamics is summarized in the spatiotemporal diagrams in Fig.~\ref{fig:spatios}, at a wind velocity slightly above the onset. Figure~\ref{fig:spatios}(a) shows that viscous solitons are emitted from the background of regular waves at moderate viscosity, but they systematically arise from a nearly sinusoidal wave packet that forms at small fetch at larger viscosity. The extent of this initial wave packet becomes smaller as $\nu_\ell$ increases: it extends over nearly ten wavelengths in Fig.~\ref{fig:spatios}(b) but shrinks to nearly two wavelengths in Fig.~\ref{fig:spatios}(d). The wavelength $\lambda$ in this wave packet, plotted as a function of $\nu_\ell$ in Fig.~\ref{fig:seuil}(b), tends towards a well defined plateau at $\lambda \simeq 21 \pm 2$ mm$~\simeq (2.2 \pm 0.2) \lambda_c$. As for the critical velocity, no dependence with the liquid depth $h$ is observed [see the inset of Fig.~\ref{fig:seuil}(b)].}

\begin{figure}[tb]
\begin{center}
\includegraphics[width=15cm]{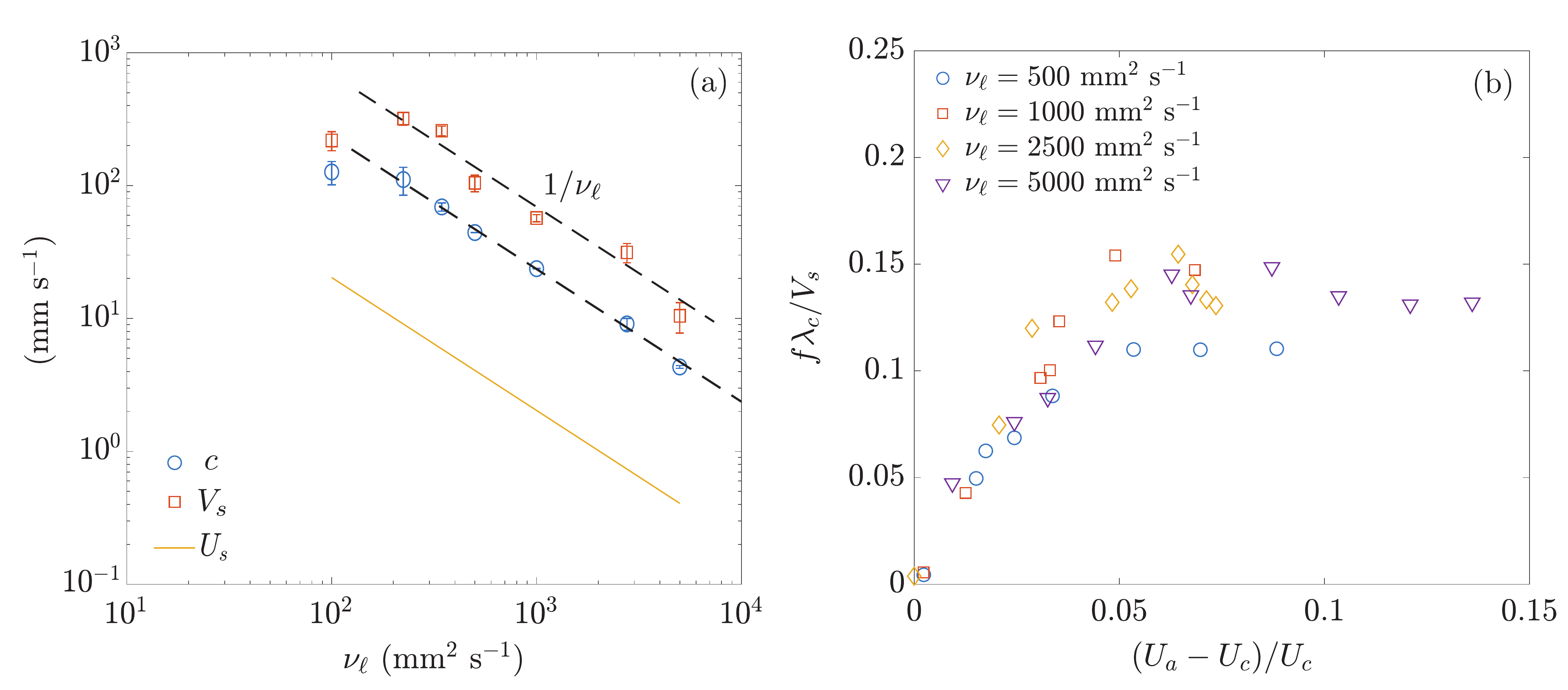}
 \caption{(a) Propagation velocity $V_s$ of solitons, phase velocity of the initial wave packet $c$ and surface drift velocity $U_s$ [computed with Eq. (\ref{eq:Us}) for a liquid depth $ h = 35$~mm] as a function of liquid viscosity $\nu_{\ell}$. Each velocity decreases as $\nu_{\ell}^{-1}$ for $\nu_\ell > 200$~mm$^2$~s$^{-1}$. (b) Normalized emission frequency of solitons,  $f \lambda_c / V_s$, as a function of the normalized wind velocity for various liquid viscosities.}
\label{fig:vitesse_frequence}
\end{center}
\end{figure}

While viscosity has no effect on the critical friction velocity and critical wavelength in the soliton regime, the dynamics of the initial wave packet and viscous solitons is strongly slowed down by the viscosity. Figure~\ref{fig:vitesse_frequence}(a) shows that both the phase velocity of the initial wave packet, $c$, and the propagation velocity of the viscous solitons, $V_s$, decrease as $\nu_\ell^{-1}$. We note that the propagation velocity $V_s$ also depends on the wind velocity (see Sec.~VII); measurements are shown here for a fixed wind velocity slightly above the onset, $U_a = 1.05 U_c$.  For this wind velocity, solitons propagate at least three times faster than the phase velocity of the initial wave packet. Note that the surface drift velocity $U_s$ (\ref{eq:Us}), although it also decreases as $\nu_\ell^{-1}$ (shown here in the case $h=35$~mm), is much smaller than both $V_s$ and $c$, suggesting that the liquid depth has a weak influence on the dynamics of the viscous solitons.

As for the propagation velocity, the mean emission frequency $f$ of solitons decreases as $\nu_\ell^{-1}$. As a consequence, the normalized emission frequency $f \lambda_c / V_s$, plotted in Fig.~\ref{fig:vitesse_frequence}(b) as a function of the normalized wind velocity $(U_a - U_c)/U_c$, collapse on a single curve independent of $\nu_\ell$. Close to the onset ($U_a < 1.05 U_c$),  the emission of solitons is erratic and the mean frequency $f \lambda_c / V_s$ increases linearly with $U_a$, while at larger velocity the emission is nearly periodic and $f \lambda_c / V_s$ tends to a plateau at $\simeq 0.13 \pm 0.02$. The minimal distance $V_s/f$ between two solitons in this asymptotic regime, $\simeq 8 \lambda_c$, is close to the total length of a soliton in the streamwise direction. This suggests that the maximum emission rate is simply governed by the time necessary for the interface to rise back to its initial level to compensate for the volume of liquid carried by the previous soliton.

\section{Relevance of the Kelvin-Helmholtz instability for the onset of the initial wave packet}

\modif{The constant critical friction velocity $u^*_c$, constant critical wavelength $\lambda$, and scaling  $c \sim \nu_\ell^{-1}$ of the phase velocity of the initial wave packet from which viscous solitons are formed, are compatible with Miles's analysis of the  Kelvin-Helmholtz instability of a viscous liquid sheared by a turbulent wind~\cite{Miles1959generation,Miles1993surface}. We examine more quantitatively in this section the agreement of our data with the theory.}

\subsection{Original and modified Kelvin-Helmholtz theories}
\label{Original KI}

We first briefly recall here the standard Kelvin-Helmholtz analysis for two inviscid non miscible fluids of different densities with a strong density ratio ($s = \rho_a / \rho_{\ell} \ll  1$)~\cite{Lamb,charru2011hydrodynamic,Rabaud_2020}. Assuming a weak wavy disturbance of the interface, the air is accelerated above the crests and decelerated above the troughs, implying pressure variations in phase with the wave profile that tends to amplify the disturbance. Gravity and surface tension define a finite velocity threshold for instability: The interface is unstable if the aerodynamic pressure overcomes the weight and surface tension. Linear stability analysis shows that the minimum wind velocity $U_c(k)$ to amplify a wave number $k$ is
\begin{equation}
U_c (k) = \frac{c_0(k)}{\sqrt{s}},
\label{eq:uck}
\end{equation}
where $c_0(k) = (g/k + \gamma k /\rho_\ell)^{1/2}$ is the phase velocity of free inviscid gravity-capillary waves. The most unstable wavelength is therefore the one that minimizes the phase velocity, i.e., the capillary wavelength $\lambda_c$, yielding the critical wind velocity
\begin{equation}
U_{KH} = \frac{c_{\min}}{\sqrt{s}},
\label{eq:UKH}
\end{equation}
with $c_{\min}$ the minimum phase velocity [Eq.~(\ref{eq:lc})]. This inviscid approach clearly fails to describe our data: for the silicon oils considered here, one has $U_{KH} \simeq 4.86$~m~s$^{-1}$ and $\lambda_c = 9.4$~mm, which is well below the experimental values ($U_c \simeq 9-10$~m~s$^{-1}$ and $\lambda \simeq 21$~mm, see Fig.~\ref{fig:seuil}).

The next step is to consider viscosity in the liquid, but still assuming a flat velocity profile in the air. This problem was first considered by Lamb~\cite{Lamb}: He derived the full dispersion relation for viscous capillary-gravity waves, and provided the appropriate boundary conditions for a purely tangential or purely normal forcing of the surface. \modif{Taylor~\cite{taylor1940generation} extended this analysis and derived the stability criterion in the case of a purely normal forcing. The key result is that the critical velocity and critical wavelength are not affected by the liquid viscosity and are still given by the inviscid Kelvin-Helmholtz theory (\ref{eq:lc})-(\ref{eq:UKH}).  However, since it ignores the effect of the wind shear stress, Taylor's analysis predicts a zero phase velocity at the onset and a scaling $c \sim \nu_\ell^{-2}$ beyond the onset~\cite{Aulnette_PhD_2021}.}

A generalization of the Kelvin-Helmholtz problem for an inviscid gas flowing with a non-constant velocity profile over a viscous liquid  was considered by Miles~\cite{Miles1959generation}. Such a non-constant velocity profile reproduces the effect of viscosity in the air, while still assuming an inviscid dynamics. This approach yields a critical wind velocity and critical wavelength larger than the flat-profile predictions (\ref{eq:lc})-(\ref{eq:UKH}).  This can be explained by the fact that large wavelengths are subjected to a stronger aerodynamic suction, because they disturb the air flow over a larger distance, where the wind is stronger. \modif{He finally refined his theory by modeling the wave-induced perturbations of the turbulent wind stresses~\cite{Miles1993surface}, from which he derives a phase velocity decreasing as $\nu_\ell^{-1}$. This scaling, different from Taylor's prediction $c \sim \nu_\ell^{-2}$, is similar to the one obtained in the Kelvin-Helmholtz-Darcy problem in the geometry of a Hele-Shaw cell~\cite{Gondret97b,plouraboue2002kelvin}.}

\subsection{A simplified version of Miles's model}
\label{sec:Miles59}

To verify to what extent the Kelvin-Helmholtz model modified by Miles describes our data, we propose in the following a simplified version of Miles's analysis. Assuming that the perturbation of the base flow in the air is irrotational, a disturbance of wavenumber $k$ modifies the flow over a distance $z \sim k^{-1}$ above the surface. The disturbance therefore experiences an aerodynamic pressure given by the air velocity at that height $z$. We can therefore modify the initial model by considering that the relevant critical wind velocity in Eq.~(\ref{eq:uck}) must be taken at a distance $z=\epsilon k^{-1}$,
\begin{equation}
u(z = \epsilon k^{-1}) = \frac{c_0(k)}{\sqrt{s}},
\label{Eq_uc_epsilon}
\end{equation}
with $\epsilon$ a non-dimensional parameter to be determined experimentally. In this simplified approach, matching the velocity profile and the rescaled dispersion relation enables the identification of $\epsilon$, and hence the critical wavenumber.

\begin{figure}[tb]
\begin{center}
\includegraphics[width=10cm]{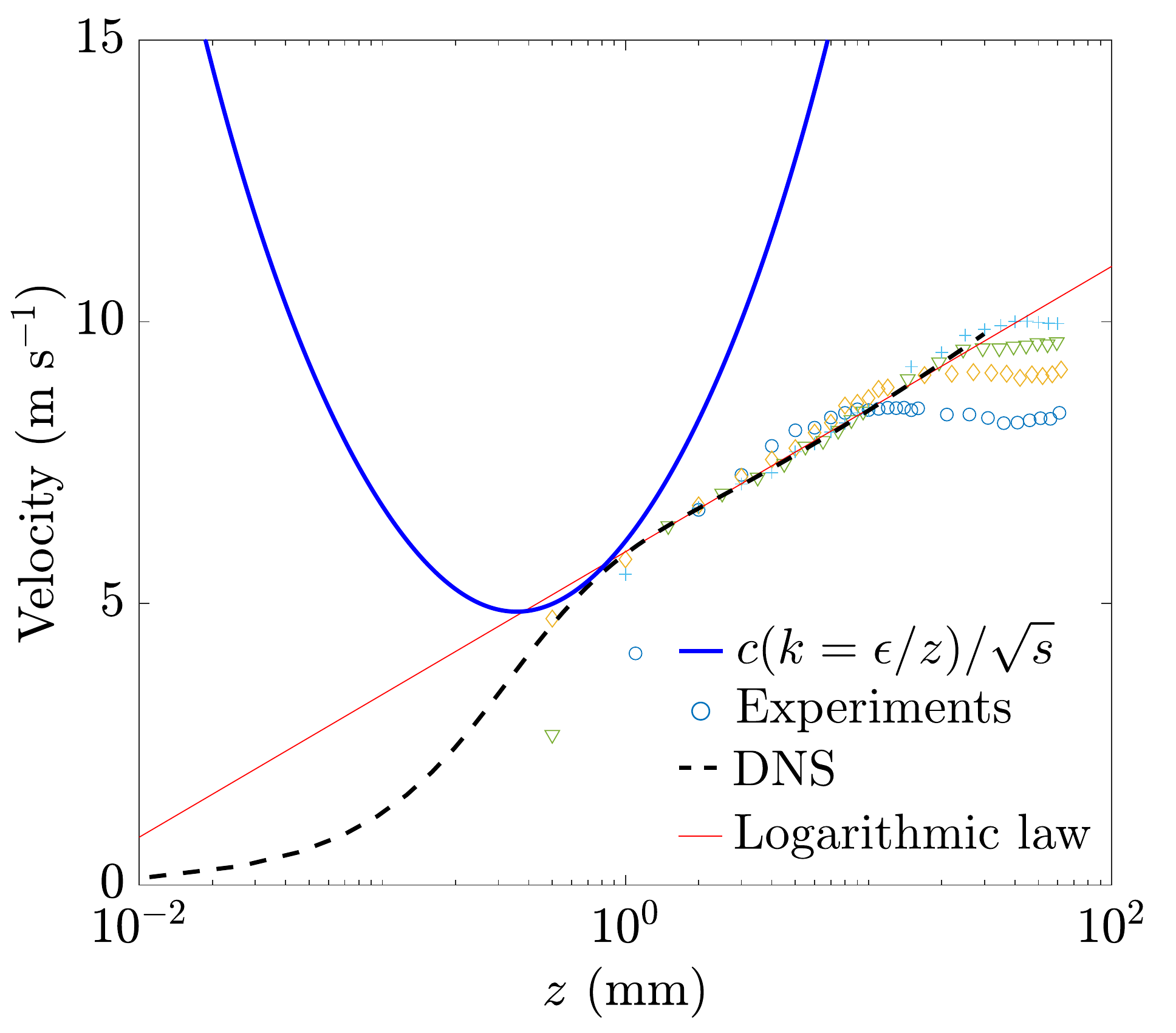}
\caption{Geometrical determination of the critical wavelength based on Eq.~(\ref{Eq_uc_epsilon}). The data shows the velocity profile measured for various fetches $x = 100, 400, 800, 1000$~mm. For each fetch the wind velocity is chosen at the onset ($U_a = 9.19, 9.58, 9.98, 10.11$~m~s$^{-1}$). Black dashed line: turbulent boundary layer profile taken from DNS~\cite{lozano2014time} for $u^*=0.44$~m~s$^{-1}$. Red line: Logarithmic approximation of the velocity profile in the inertial subrange (\ref{eq:loglaw}). Blue line: Rescaled dispersion relation $c(k=\epsilon/z) / \sqrt{s}$,  for a value $\epsilon = 0.23$ chosen such that it intersects the DNS velocity profile.}
\label{fig:Miles59}
\end{center}
\end{figure}

We plot in Fig.~\ref{fig:Miles59} the velocity profiles $u(z)$ measured at different fetches $x$ for a wind velocity $U_a$ corresponding to the onset of waves. As $x$ is increased, the saturation value of $u(z)$ at large $z$ and the boundary layer thickness $\delta$ increase, in agreement with Fig.~\ref{fig:membrane}, but the data approximately collapse in the inertial subrange, for $z \ll \delta$. Since the velocity profile in the inertial subrange is governed by the friction velocity $u^*$, this collapse is a confirmation of the constant $u_c^* = 0.44$~m~s$^{-1}$ at the onset of waves. Because of the lack of reliable data for $z<1$~mm, we extrapolate our measurements using the velocity profile obtained from direct numerical simulation (DNS) of a turbulent channel flow~\cite{lozano2014time} scaled for $u^* = 0.44$~m~s$^{-1}$ (black dashed line). In the inertial subrange (1~mm~$<z<10$~mm), the experimental and DNS data coincide, and are approximately described by the classical logarithmic law
\begin{equation}
u(z) = u^* [\kappa^{-1} \log(z/ \delta_\nu) + C^+],
\label{eq:loglaw}
\end{equation}
with $\delta_\nu = \nu_a / u^* \simeq 35$~$\mu$m the viscous sublayer thickness, $\kappa = 0.41$ the von K\'arm\'an constant, and $C^+ = 5$. To determine $\epsilon$, we apply a horizontal shift to the rescaled dispersion relation, $c(k = \epsilon z^{-1}) / \sqrt{s}$, until it becomes tangent to the DNS profile $u(z)$.
The intersection occurs for the value $\epsilon = 0.23 \pm 0.01$, at a very small distance $z \simeq 0.7$~mm above the surface. This critical distance falls in the buffer layer ($z \simeq 20~\delta_\nu$)  between the viscous and inertial subranges. In this region our velocity measurements are no longer reliable, which justifies using the extrapolation with the DNS data. The resulting critical wavelength, $\lambda = 2\pi z / \epsilon \simeq (2.1 \pm 0.4) \lambda_c$, is in good agreement with the measured value, $\lambda \simeq (2.2 \pm 0.2) \lambda_c$, showing that this simplified method produces a correct description of the instability.
 
We note that the critical distance $z \simeq 0.7$~mm  is much smaller than the boundary layer thickness $\delta \simeq 10-30$~mm, suggesting that the outer structure of the boundary layer does not affect the critical wavelength. This is consistent with the observation that as $x$ is increased, the boundary layer thickens [see Fig.~\ref{fig:membrane}(a)] but the critical wavelength remains constant. We also note that with a critical distance in the buffer layer ($z \simeq 20 \delta_\nu$), the logarithmic approximation (\ref{eq:loglaw}), which was used in Miles's analysis~\cite{Miles1959generation}, no longer applies, so the exact velocity profile is necessary to determine the critical wavelength: solving Eq.~(\ref{Eq_uc_epsilon}) with the logarithmic approximation (\ref{eq:loglaw}) would yield a smaller critical wavelength, not compatible with our data.

\modif{We finally turn to the phase velocity of the initial wave packet. In his refined analysis of the Kelvin-Helmholtz instability, Miles~\cite{Miles1993surface} derives the phase velocity as
\begin{equation}
c = \frac{s^{1/2} \kappa u^*}{k \nu_\ell} c_0,
\label{eq:cnum}
\end{equation}
with $c_0$ the inviscid phase velocity. The scaling $\nu_\ell^{-1}$ is in good agreement with our measurements in Fig.~\ref{fig:vitesse_frequence}(a), although the numerical factor is approximately five times smaller. This discrepancy may be related to an inaccurate modeling of the Reynolds stress fluctuations in Miles's theory.  We also note that the predicted phase velocity (\ref{eq:cnum}) is of the order of the surface drift velocity, $U_s$ [Eq.~(\ref{eq:Us})], plotted in Fig.~\ref{fig:vitesse_frequence}(a) for a liquid depth $h=35$~mm. Although this is not enough to explain the factor 5, this suggests that the sheared profile in the liquid may affect the instability and modify the wave properties.}

\section{Transition to the viscous soliton regime}

We turn back to the dependence of the friction velocity threshold with viscosity [Fig.~\ref{fig:seuil}(a)], and address here the question of the transition between the $u_c^* \sim\nu_\ell^{1/5}$ regime at low viscosity and the $u_c^* \sim\nu_\ell^{0}$ regime at large viscosity. In the present experiments performed in silicon oils, the crossover between these  two regimes occurs for a critical viscosity $\nu_{\ell,c} \simeq 200 \pm 40$~mm$^2$~s$^{-1}$, a value slightly larger than the one reported by Paquier {\it et al.}~\cite{Paquier_2016}, $\nu_{\ell,c} \simeq 100 \pm 50$~mm$^2$~s$^{-1}$, in experiments performed in aqueous solutions with larger surface tension and density. This raises the question of the relevant physical parameters that govern the critical viscosity between the two regimes.

Following the Kelvin-Helmholtz prediction (\ref{eq:UKH}), we can write the constant critical friction velocity at large viscosity in the form
\begin{equation}
u^*_{c,KH} = A \frac{c_{min}}{\sqrt{s}},
\label{eq:uckh}
\end{equation}
with $A$ a parameter related to the details of the velocity profile. From the critical friction velocity $u_c^* \simeq 0.44\pm 0.01$~m~s$^{-1}$ found in our experiment, we have $A = 0.09 \pm 0.002$. For $\nu_{\ell} < 200 \pm 40$~mm$^2$~s$^{-1}$, the smaller critical friction velocity indicates that a more unstable mechanism is at play. This low-viscosity instability was identified as the transition between wrinkles (the linear surface response to the turbulent pressure fluctuations traveling in the boundary layer) and regular waves~\cite{Paquier_2016,Perrard_2019}. The wrinkle amplitude results from the balance between the energy supplied by the pressure fluctuations and the viscous dissipation in the liquid,
$$
\zeta_{rms} \simeq C s \, \delta \left( \frac{ u^{*3}}{g \nu_\ell} \right)^{1/2},
$$
with $C \simeq 0.02$. Assuming that the regular waves are triggered when the wrinkle amplitude becomes larger than a fraction of the viscous sublayer thickness $\delta_\nu = \nu_a / u^*$ yields the critical friction velocity
\begin{equation}
u^*_{c,RW} = K s^{-2/5} \left( \frac{g \nu_\ell \nu_a^2}{\delta^2}
\label{eq:ucrw}
\right)^{1/5},
\end{equation}
where $K\simeq 1.6\pm 0.2$ is an empirical constant fitting the data in Fig.~\ref{fig:seuil}(a). 
The transition between regimes (\ref{eq:uckh}) and (\ref{eq:ucrw}) is therefore expected for the critical viscosity
\begin{equation}
\nu_{\ell,c}  \simeq (A/K)^5 s^{-1/2} \frac{\delta^2 c_{min}^5}{g \nu_a^2}.
\label{eq:nulc1}    
\end{equation}
This critical viscosity therefore depends both on the liquid properties and the turbulent boundary layer flow in the air. Testing the validity of this crossover would require to vary the fluid properties over a wide range, which is difficult using conventional laboratory fluids.

It is interesting to compare this critical viscosity to the viscosity for which the initial wave packet becomes critically damped. Since the wavelength in the initial wavepacket is close to the capillary length ($\lambda \simeq 2.2 \lambda_c$), this crossover viscosity corresponds to a Reynolds number $Re = \lambda_c c_{min}/\nu_\ell$ of the order of unity or, equivalently, to a Morton number
\begin{equation}
Mo = \frac{g \rho_\ell^3 \nu_\ell^4}{\gamma^3}
\label{eq:Morton}    
\end{equation}
of the order of unity (with $Mo \simeq Re^{-4}$). This number is a defining property of the liquid: \modif{for $Mo < O(1)$ waves at $\lambda_c$ are propagating, while for $Mo > O(1)$ they are over-damped.}

\modif{In our experiments, the Morton number at the transition viscosity $\nu_{\ell,c} \simeq 200 \pm 40$~mm$^2$~s$^{-1}$ is $Mo \simeq 2 \pm 1$. This indicates that the waves in the Kelvin-Helmholtz regime satisfy $Mo > 2$, i.e. they fall in the overdamped regime. This may be related to the subcritical nature of the transition to viscous solitons:} because of the vanishing energy propagation velocity in an overdamped wave, the energy supplied by the air flow cannot be radiated and may amplify the surface deformations. This implies that Kelvin-Helmholtz waves for liquids with small Morton number (e.g., for larger surface tension) could behave differently,  possibly lacking the generation of viscous solitons. Here again, identifying fluids allowing the Morton number to be varied at the transition to the Kelvin-Helmholtz regime is difficult.

% --------------------------------------------------------------------------
\section{Mechanical excitation of viscous solitons}

\begin{figure}[tb]
\begin{center}
\includegraphics[width=15cm]{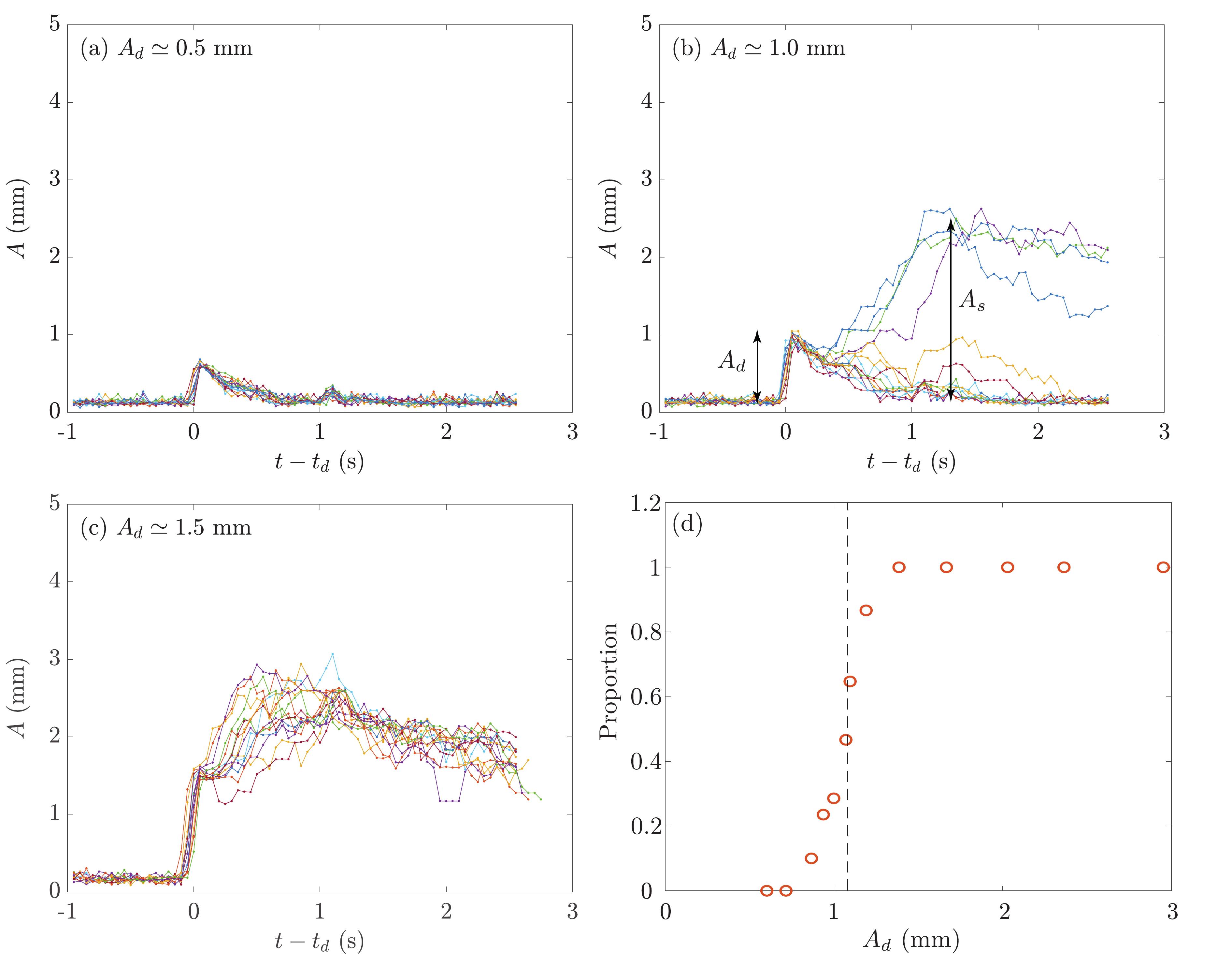}
\caption{Surface deformation amplitude as a function of time for 15 wavemaker strokes for different values of the deformation amplitude, for $U_a= 8.4$ m~s$^{-1}$ and $\nu = 1000$ mm$^2$~s$^{-1}$: (a) $A_d = 0.5$~mm (no soliton produced); (b) $A_0 = 1$ mm (some strokes produce solitons) and (c) $A_0 = 1.5$~mm (solitons produced all the time).  (d) Fraction of solitons generated over 15 strokes as a function of the deformation amplitude $A_d$. The dashed vertical line shows the deformation amplitude for which 50\% of the wavemaker strokes produce a soliton.}
\label{fig:proba_SC}
\end{center}
\end{figure}

The subcritical nature of the instability producing viscous solitons manifests itself through the spatial decay of the shear stress in a developing boundary layer: a soliton appears when the local shear stress reaches the upper threshold $\tau_1 = \rho_a u^{*2}_c$, and continues its course as long as the local shear stress remains larger than a second lower threshold $\tau_2 < \tau_1$. In the hysteresis window $[\tau_2, \tau_1]$, both the flat surface and the soliton are stable solutions, separated in principle by an unstable branch, which represents the minimum surface deformation required to \modif{bring} the system from the flat surface branch to the soliton branch. While reconstructing the stable soliton branch of the bifurcation diagram is possible using the natural evolution of the solitons amplitude along their course~\cite{Aulnette_2019}, exploring the unstable branch requires applying a finite amplitude disturbance to force the system to the soliton branch.

To determine this unstable branch, we have set up a wavemaker made of a horizontal cylinder, $3$~mm in diameter, \modif{located at the depth $z_c = -15$~mm below the free surface at a fetch $x=300$~mm. We apply a vertical stroke of amplitude $A_0$ between 0 and 4~mm, resulting in a small surface deformation of amplitude $A_d \leq 3$~mm. The dependence of the surface deformation $A_d$ with the wavemaker amplitude $A_0$, wavemaker velocity and immersion depth is complex (see details in Ref.~\cite{Aulnette_PhD_2021}), but we found that the generation of solitons is essentially governed by the resulting surface deformation $A_d$, so we simply use $A_d$ as the control parameter in the following.}

\begin{figure}[tb]
\begin{center}
\includegraphics[width=9cm]{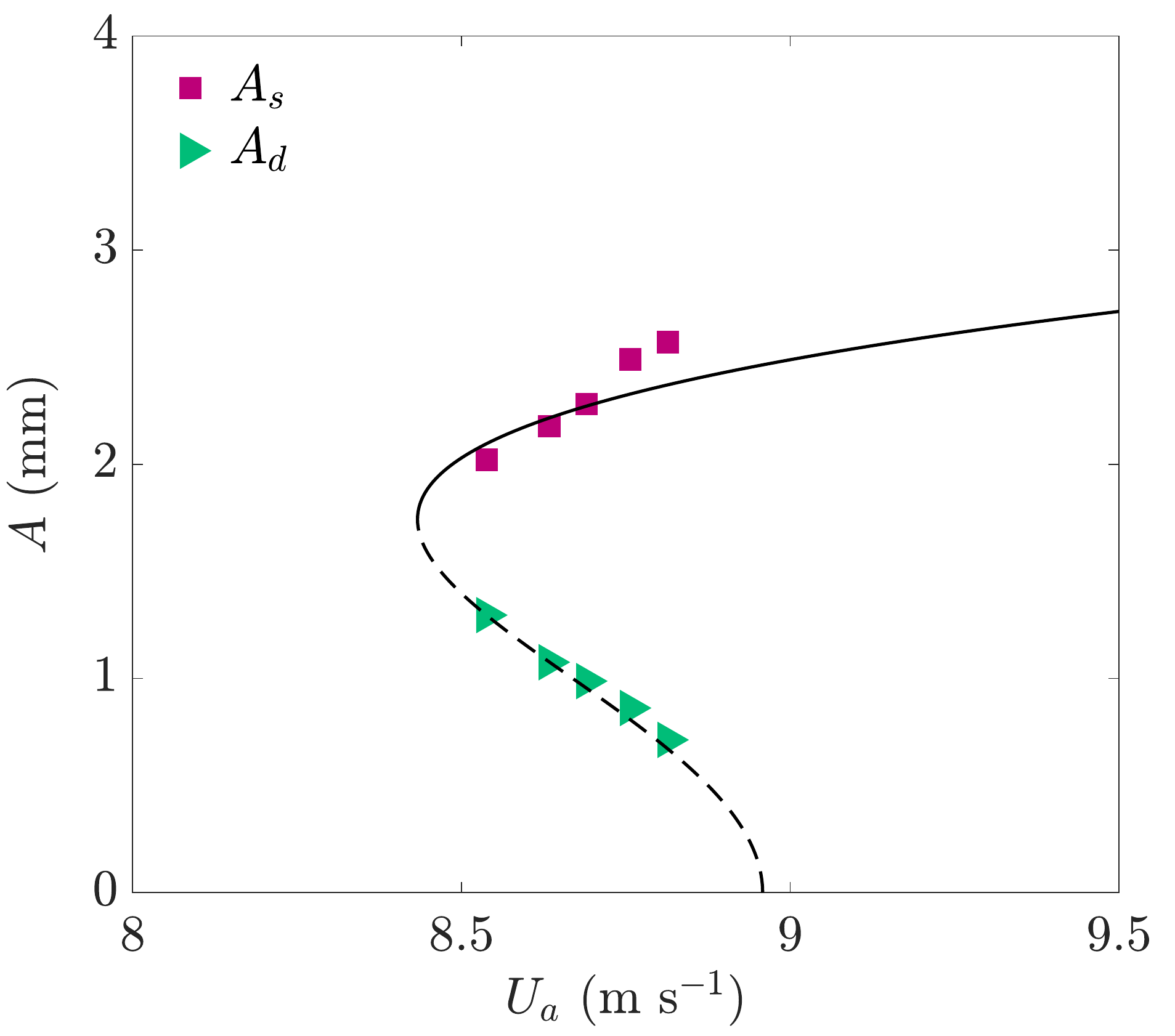}
\caption{Subcritical bifurcation diagram reconstructed from the wave-maker experiments ($\nu_\ell = 1000$ mm$^2$~s$^{-1}$). Green triangles: Minimum surface deformation $A_d$ to produce solitons with probability 50\%. Red squares: resulting soliton amplitude $A_s$. The line shows a fit with Eq.~(\ref{eq:subua}).}
\label{fig:diagramme_bifurcation}
\end{center}
\end{figure}

The results are illustrated in Figs.~\ref{fig:proba_SC}(a)-(c), showing the time evolution of the surface amplitude right after the wavemaker for 15 strokes at a given wind slightly below the onset, $U_a = 0.97 U_c$, for different forcing amplitudes $A_d$.  Since the wavemaker is located at a single fetch $x$, this wind velocity corresponds to a fixed friction velocity slightly below the onset $u_c^*$. For small $A_d$ none of the wave-maker strokes produce a soliton, while for large $A_d$ they systematically produce solitons. Interestingly, for an intermediate range of  $A_d$, solitons are generated only with a certain probability: this is a manifestation of the stochastic emission of solitons close to the unstable branch. The fraction of wavemaker strokes producing a soliton, plotted in Fig.~\ref{fig:proba_SC}(d), increases gradually from 0 to 1 as the disturbance amplitude is increased. This smooth transition suggests that producing a soliton not only depends on the disturbance amplitude, but also on the value of the fluctuating airflow at the instant of the perturbation. 

To reconstruct the unstable branch of the bifurcation diagram, we  define the critical amplitude $A_d$ such that solitons are generated with a probability of 50\%. We plot this critical amplitude $A_d$, as well as the resulting soliton amplitude $A_s$, in Fig.~\ref{fig:diagramme_bifurcation} as functions of the wind velocity $U_a$. As expected for a subcritical bifurcation, the closer the wind velocity $U_a$ to the natural onset $U_{c}$, the smaller the surface disturbance to trigger a soliton. Our data is well described by the normal form of a subcritical bifurcation,
\begin{equation}
U_a (A) = U_c - a A^2 + b A^4,
\label{eq:subua}
\end{equation}
with $U_c = 8.95$~ms$^{-1}$ (corresponding to $u^*_c = 0.44$~m~s$^{-1}$ for this fetch), $a = 0.35$~m$^{-1}$s$^{-1}$, and $b = 0.06$~m$^{-3}$s$^{-1}$, providing a clear confirmation of the subcritical nature of the instability.

The stochastic triggering of solitons observed here is in good agreement with the large distribution of waiting times between solitons observed close to the natural onset $U_c$ without mechanical forcing. Assuming that the Kelvin-Helmholtz waves provide the seed disturbance from which the viscous solitons grow, small turbulent fluctuations bring the system stochastically on both sides of $U_c$, alternately amplifying and attenuating the surface deformation, resulting in unpredictable generation of solitons. This stochasticity disappears at larger wind velocity, for which any disturbance of vanishing amplitude systematically \modif{produces} solitons.

%---------------------------------------------------------------------------------------------------------------------
\section{A model for the propagation velocity} \label{sec:NL_dynamics}

\begin{figure}[tb]
\begin{center}
\includegraphics[width=10cm]{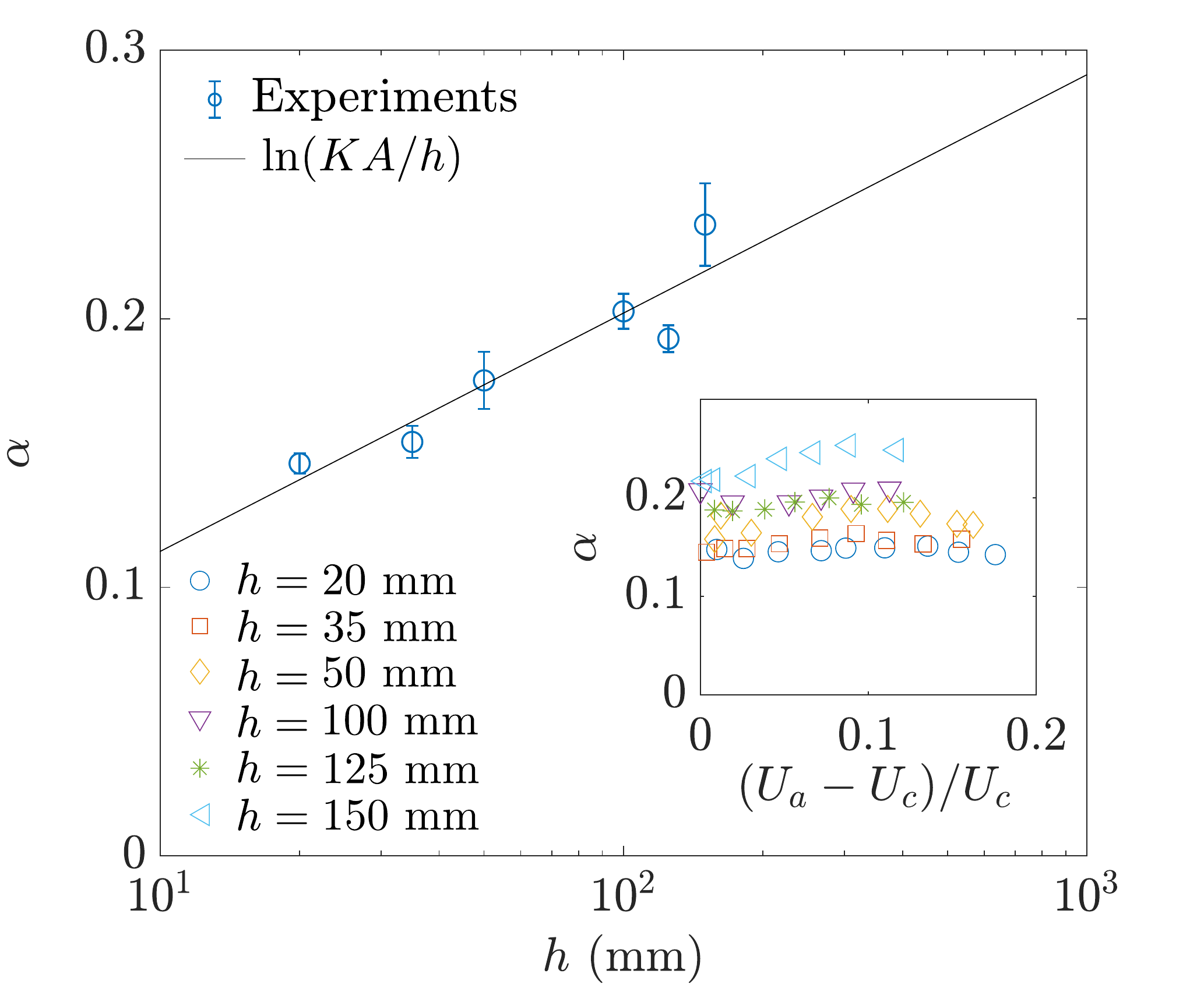}
\caption{Normalized soliton velocity $\alpha$ (\ref{eq:Vs_modele}) as a function of liquid depth $h$ for $\nu_\ell = 1000$~mm$^2$~s$^{-1}$. The line shows the best fit with the logarithmic model (\ref{eq:lnkah}). Inset: $\alpha$ as a function of wind velocity for different liquid depths.}
\label{fig:alpha_vs_h}
\end{center}
\end{figure}

We finally address the dependence with the various flow parameters of the propagation velocity of viscous solitons. With a Reynolds number based on the soliton size and velocity $Re_s = V_s A /\nu_\ell < 5$ for our liquid viscosities ($\nu_\ell = 200-5000$~m$^2$s$^{-1}$), the propagation of solitons is controlled by the viscous diffusion in the liquid of the momentum supplied by the wind. Noting $\rho_a U_a^2 A$ the order of magnitude of the aerodynamic drag and $\rho_\ell \nu_\ell V_s$ the viscous friction in the liquid (both per unit of transverse length), dimensional analysis implies a soliton velocity in the form
\begin{equation}
V_s \simeq \alpha s \frac{U_a^2 A}{\nu_\ell},
\label{eq:Vs_modele}
\end{equation}
with $\alpha$ a numerical coefficient that may depend both on the nature of the air flow above the soliton and the liquid flow below it. The scaling with $U_a$ and $A$ was already confirmed in the experiments of Aulnette {\it et al.}~\cite{Aulnette_2019} performed at a single viscosity $\nu_\ell \simeq 1000$~mm$^2$~s$^{-1}$, whereas the scaling with viscosity is confirmed by the present experiments: Figure~5(a) shows that $V_s \propto \nu_\ell^{-1}$ at a fixed wind velocity above the onset, $U_a \simeq 1.05 U_c$. It remains, however, a weak dependence in the liquid depth that remains to be modeled: Figure~\ref{fig:alpha_vs_h} shows that the factor $\alpha$ in Eq.~(\ref{eq:Vs_modele}) is indeed independent of $U_a$, but it slightly increases from 0.14 to 0.22 as $h$ is increased.  We analyze here the flow in the liquid induced by the viscous soliton, and propose a model for this dependence of $\alpha$ with $h$. 

\begin{figure}[tb]
\begin{center}
\includegraphics[width=15cm]{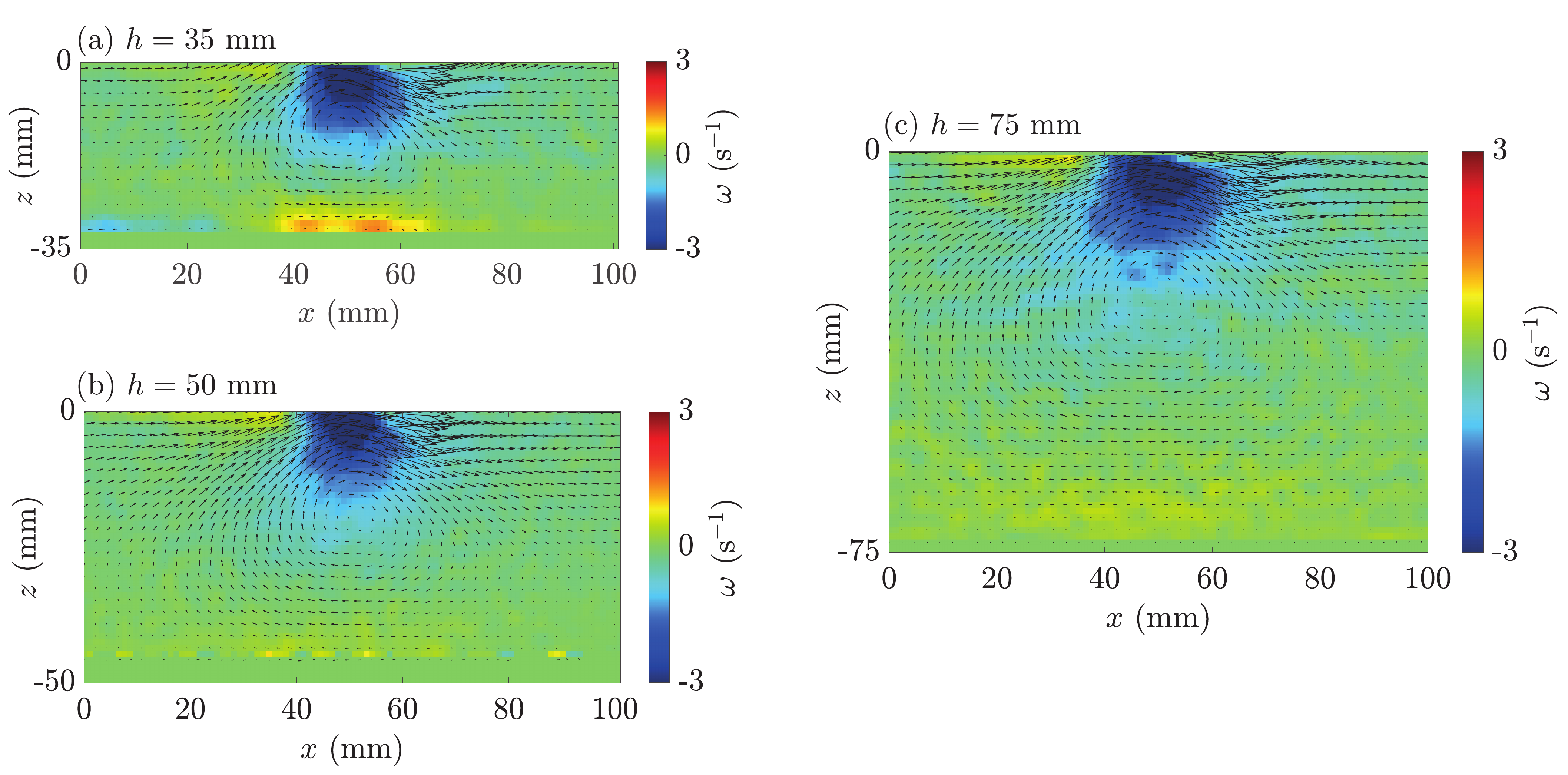}
\caption{Flow below a soliton for different liquid depths (a) $h = 35$ mm, (b) $h = 50$ mm, and (c) $h = 75$ mm (wind velocity $U_a = 9.15$ m~s$^{-1}$). The vorticity field shows an approximately isotropic region under the soliton, of size independent of $h$.}
\label{fig:PIV}
\end{center}
\end{figure}

\begin{figure}[tb]
\begin{center}
\includegraphics[width=15cm]{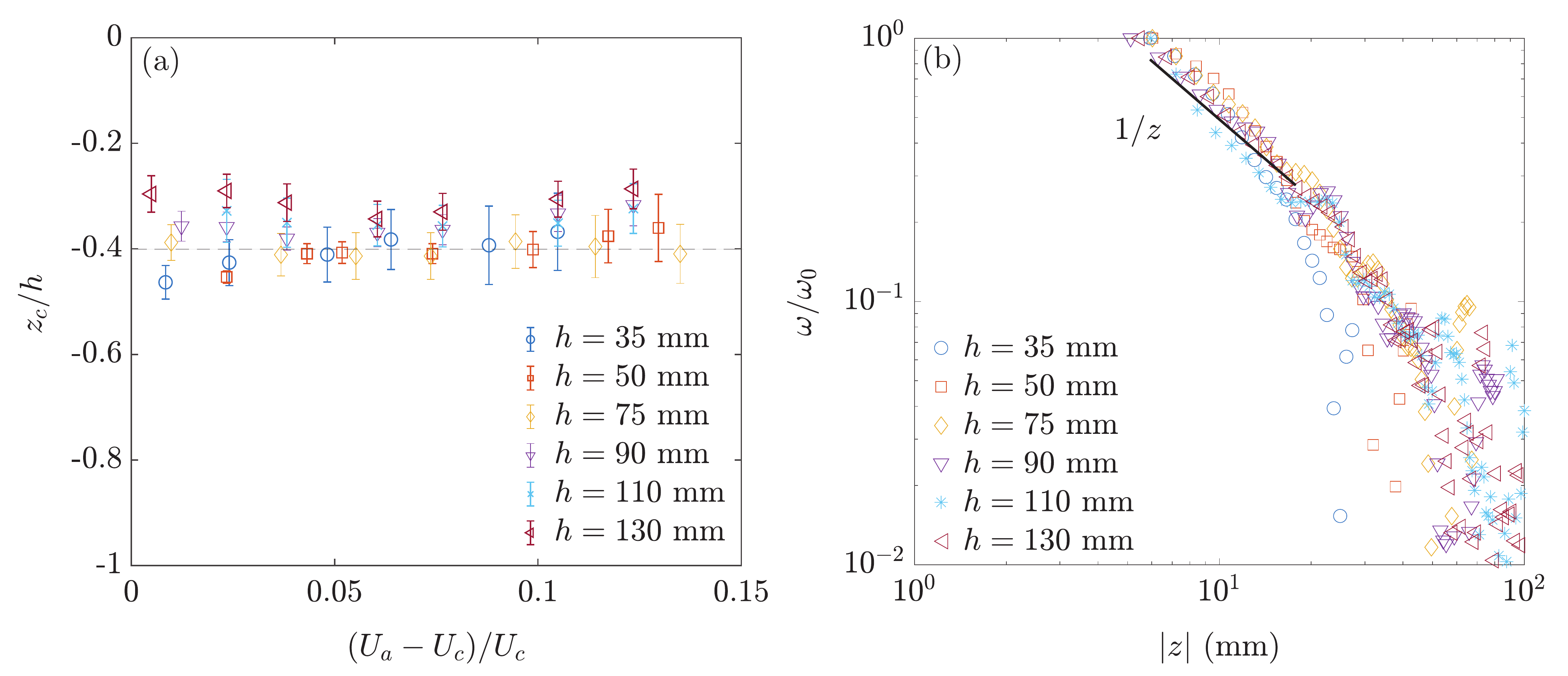}
\caption{(a) Vortex location $z_c$ as a function of the wind velocity $U_a$ for several liquid depths $h$; the average is $z_c \simeq -0.4 h$. (b) Mean vorticity profile $\omega(z)$ under the soliton, normalized with the maximum vorticity close to the interface, for several liquid depths for $U_a = 1.05 U_c$.  The vorticity profiles are averaged over a large set of solitons.}
\label{fig:profondeur_PIV}
\end{center}
\end{figure}

PIV measurements in the liquid under a soliton are shown in Fig.~\ref{fig:PIV} for various liquid depths for $\nu_\ell = 1000$~mm$^2$~s$^{-1}$ (the flow inside the soliton cannot be measured due to the difficult optical access). The flow consists in an approximately isotropic vorticity patch located under the soliton, associated with a vortex flow that extends to the bottom of the liquid layer. Interestingly, the depth of zero velocity, marking the center of the vortex, is governed by the liquid layer thickness $h$, with $z/h \simeq -0.4$ [see Fig.~\ref{fig:profondeur_PIV}(a)], whereas the size of the vorticity patch does not vary with $h$.

These properties are characteristic of a two-dimensional Stokes flow induced by a disturbance of size much smaller than the domain: vorticity diffuses isotropically far from the disturbance, while the velocity field is affected by the boundary conditions.  In this problem, vorticity is the solution of the stationary Stokes equation $\nabla^2 \omega=0$ satisfying the boundary condition at the surface given by the applied shear stress, $\omega(x, z=0) = - \tau(x) / \eta_\ell$. In the absence of soliton, the shear stress is approximately uniform in $x$, and the solution to the Stokes problem reduces to the classical Poiseuille flow, with a linearly decreasing vorticity $\omega(z) = \omega_0 (1+3z/2h)$ with $\omega_0 = -\tau/\eta_\ell$ and a parabolic velocity profile. Modeling a viscous soliton as a two-dimensional (2D) localized shear stress translating at constant velocity, $\tau(x-V_s t)$, the solution to the stationary Stokes problem is a 2D Stokeslet following the source. By linearity, the flow is the sum of the Poiseuille contribution due the uniform wind shear stress and the Stokeslet contribution due to the localized stress disturbance. The soliton velocity being at least one order of magnitude larger than the surface drift velocity [see fig.~\ref{fig:vitesse_frequence}(a)], this suggests that the Stokeslet contribution is much larger than the Poiseuille contribution.

In a semi-infinite domain, the far-field vorticity and horizontal velocity profiles in the frame of the traveling shear stress source are given by the 2D Stokeslet solution~\cite{Lamb,Pozrikidis1992},
\begin{equation}
\omega (X=0, z) = \frac{F_0}{\pi \eta_{\ell} z}, \qquad u_x (X=0,z) = - \frac{F_0}{2 \pi \eta_{\ell}} \ln \left(\frac{z}{z_0} \right),
\label{eq:stokeslet2d}    
\end{equation}
with $X=x-V_s t$, $F_0 = \int \tau(x) dx$ the force per unit of transverse length of the source and $z_0$ the depth at which velocity changes sign (vortex center). Vorticity decays away from the source, while the velocity diverges logarithmically, a generic feature of two-dimensional diffusion in an unbounded domain.

The vorticity profiles measured by PIV shown in Fig.~\ref{fig:profondeur_PIV}(b) confirm the $|z|^{-1}$ decay for an intermediate range of $|z|$. This indicates that the semi-infinite 2D Stokeslet model provides a good description of the flow under a soliton far from the surface and from the bottom of the tank. The value of $z_0$, which is not prescribed in an unbounded domain, must be fixed by mass conservation: it must therefore scale as $h$, in good agreement with the normalized constant value of the vortex center, $z_c/h \simeq -0.4$; note that this value is close to the expected zero-crossing at $-1/3$ in the Poiseuille flow.

Since the problem is quasi-stationary, up to now the propagation velocity $V_s$ is a free parameter of the model. Its value must result from the dynamics of the Stokeslet itself: the source generates a flow that depends on the liquid depth, which in turn induces a surface drift that transports the source. We can therefore estimate $V_s$ as the surface velocity induced by the shear stress applied to the soliton. This surface velocity cannot be computed directly from Eq.~(\ref{eq:stokeslet2d}), which is valid only in the far field. We must therefore introduce a cutoff, and take for $V_s$ the value of $u_x$ at a depth $z$ of the order of the soliton size $A$. Taking $z_0 \propto h$ for the vortex center and $F_0 \propto \rho_a U_a^2 A$ for the Stokeslet force per unit of transverse length, we obtain a propagation velocity in the form
$$
V_s \propto s \frac{U_a^2 A}{\nu_\ell} \ln (K A/h),
$$
with $K$ a numerical factor. Accordingly, the prefactor in Eq.~(\ref{eq:Vs_modele}) writes
\begin{equation}
\alpha = \ln(K A/h),    
\label{eq:lnkah}
\end{equation}
which provides a good description of our measurements in Fig.~\ref{fig:alpha_vs_h} with $K \simeq 3$. This confirms that the soliton velocity results from a balance between the aerodynamic drag and the viscous drag with a weak dependence in the liquid depth.

The logarithmic dependence in liquid depth is reminiscent of the Stokes drag for a two-dimensional object in an infinite domain, $F \propto \eta_\ell V / \ln(1/Re)$, with $Re$ the Reynolds number based on the disturbance size and velocity~\cite{Lamb}. Since this model assumes that the flow is governed by the viscous diffusion across the liquid layer, it holds only for a liquid depth smaller than the diffusion lengthscale $\nu_\ell / V_s$, i.e. for a small Reynolds number $Re_h = V_s h / \nu_l$. The good agreement of our data with Eq.~(\ref{eq:lnkah}) suggests that this is indeed the case here, at least up to $Re_h \simeq 10$.

%---------------------------------------------------------------------------------------------------------------------
\section{Conclusion} 

In this paper, we investigated experimentally the influence of liquid viscosity and depth on the dynamics of viscous solitons. The main findings of Aulnette {\it et al.}~\cite{Aulnette_2019} are confirmed by the present experiments: Viscous solitons are sub-critically generated when a liquid of sufficient viscosity is sheared by a turbulent wind, and propagate downwind at a velocity that results from the balance between aerodynamic drag and viscous friction. They arise from an initial wave packet that forms at small fetch, where the wind shear stress is larger. \modif{The properties of this initial wave packet are in good agreement with Miles's analysis of the Kelvin-Helmholtz instability for a highly viscous fluid sheared by a turbulent wind~\cite{Miles1959generation,Miles1993surface}: the critical friction velocity and critical wavelength are independent of viscosity, while the phase velocity decreases as $1/\nu_\ell$.}

Varying the liquid viscosity over a wide range allowed us to specify the mechanism governing the propagation of viscous solitons, refining the model proposed in Ref.~\cite{Aulnette_2019} by including the effect of the liquid depth.  Our PIV measurements show that the flow in the liquid is well described by a 2D Stokeslet flow, indicating that the friction essentially results from the diffusion of vorticity induced by a localized 2D shear stress disturbance. The resulting propagation velocity shows a  logarithmic dependence in liquid depth, which is a direct consequence of a diffusion process in a bounded domain.

Finally, we performed experiments to explore the subcritical nature of the generation of viscous solitons. When spontaneously forced by the wind, viscous solitons are generated in a region of large shear stress but, once formed, they are sustained by the wind and propagate in a region of lower stress. Using a mechanical forcing, we measured the minimum disturbance necessary to trigger a soliton, allowing us to reconstruct the unstable branch of the subcritical bifurcation diagram. These experiments highlight the stochastic nature of the instability below the critical friction velocity: generating a viscous \modif{soliton} requires the combination of a disturbance of sufficient amplitude (provided either by a natural unstable wave packet or by the mechanical forcing) and favorable instantaneous turbulent fluctuations in the air flow. 

Our study rises a number of questions about the transition between weakly nonlinear waves at small viscosity and strongly nonlinear viscous solitons at larger viscosity. For intermediate viscosities, $\nu_\ell \simeq 50-200$~mm$^2$~s$^{-1}$, a complex interplay between viscous solitons and weakly nonlinear waves takes place. Our observations suggests that, in that intermediate regime, viscous solitons are triggered by regular waves of sufficient amplitude, whereas they are triggered by Kelvin-Helmholtz waves at larger viscosity. This transition between competing instabilities producing viscous solitons goes beyond the present study and \modif{need} further investigations.

\begin{acknowledgments}

We are grateful to F. Charru, G. Dietze, P. Gondret, W. Herreman and J. Magnaudet  for fruitful discussions. We thank  A. Aubertin, L. Auffray, J. Amarni, N. Lanchon and R. Pidoux for experimental help. This work was supported by the project ``ViscousWindWaves'' (Project No. ANR-18-CE30-0003) of the French National Research Agency.

\end{acknowledgments}

%%%%%%%%%%%%%
%\bibliographystyle{plain}
\bibliographystyle{unsrt}
\bibliography{Biblio_solitons}
%%%%%%%%%%%%%

\end{document}